\documentclass[a4paper,11pt,twoside]{article}

\usepackage{amsmath}
\newcommand{\lap}[1]%
        {\boldsymbol{\lambda}_{\sect{#1}}^{\phantom{\dagger}}}

\newcommand{\tlad}[1]%
        {\tilde{\boldsymbol{\lambda}}_{\sect{#1}}^\dagger}

\newcommand{\ord}[1]{\mathcal{O}\left( #1 \right)}

\newcommand{\Vev}[1]{\left\langle #1\right\rangle}

\newcommand{\sect}[1]{{\mathnormal{#1}}}

\newcommand{\nohyphens}%
        {\hyphenpenalty=10000\exhyphenpenalty=10000\relax}

\makeatletter
\renewcommand{\fnum@table}{\textbf{\tablename~\thetable}}
\renewcommand{\fnum@figure}{\textbf{\figurename~\thefigure}}
\makeatother

\newcommand{\myappendix}%
        {\appendix}

\newcounter{myenumi}

\renewcommand{\themyenumi}{\roman{myenumi}}
{\end{list}}

\makeatletter
\if@twoside
  \renewcommand{\ps@headings}{%
      \let\@oddfoot\@empty\let\@evenfoot\@empty         
      \def\@evenhead{\thepage\hfil\bfseries\leftmark}   
      \def\@oddhead{{\bfseries\rightmark}\hfil\thepage} 
      \let\@mkboth\markboth
    \def\chaptermark##1{%
      \markboth {%
        \ifnum \c@secnumdepth >\m@ne
            \@chapapp\ \thechapter. \ %
        \fi
        ##1}{}}%
    \def\sectionmark##1{%
      \markright {%
        \ifnum \c@secnumdepth >\z@
          \thesection. \ %
        \fi
        ##1}}}
\else
  \renewcommand{\ps@headings}{%
    \let\@oddfoot\@empty                               
    \def\@oddhead{{\bfseries\rightmark}\hfil\thepage}  
    \let\@mkboth\markboth
    \def\chaptermark##1{%
      \markright {%
        \ifnum \c@secnumdepth >\m@ne
            \@chapapp\ \thechapter. \ %
        \fi
        ##1}}}
\fi
\makeatother

\newcommand{\ppb}{\phi\bar{\phi}}
\newcommand{\ppd}{\phi\phi^\dagger}
\newcommand{\bppd}{\bar{\phi}\bar{\phi}^\dagger}
\newcommand{\dppb}{\phi^\dagger\bar{\phi}^\dagger}
\newcommand{\aab}{A\bar{A}}
\newcommand{\aad}{AA^\dagger}
\newcommand{\baad}{\bar{A}\bar{A}^\dagger}
\newcommand{\daab}{A^\dagger\bar{A}^\dagger}
\newcommand{\pa}{\bar{\phi}A\phi^\dagger}
\newcommand{\pab}{\phi\bar{A}\bar{\phi}^\dagger}
\newcommand{\pad}{\bar{\phi}^\dagger A^\dagger \phi}
\newcommand{\pabd}{\phi^\dagger\bar{A}^\dagger\bar{\phi}}

\newcommand{\fracwithdelims}[4]{\left#1 \frac{#3}{#4} \right#2}

\newlength{\myem}
\newcommand{\sep}[1]{\makebox[\myem][c]{#1}}
\newcounter{mysubequation}[equation]
\renewcommand{\themysubequation}{\alph{mysubequation}}
\newcommand{\mytag}{\stepcounter{mysubequation}%
\tag{\theequation\protect\sep{\themysubequation}}}
\newcommand{\globallabel}[1]{\refstepcounter{equation}\label{#1}}

\makeatletter
\renewcommand{\section}{\@startsection{section}{1}{0em}{-\baselineskip}%
{\baselineskip}{\normalfont\large\bfseries}}
\renewcommand{\subsection}%
{\@startsection{subsection}{2}{0em}{-0.7\baselineskip}%
{0.7\baselineskip}{\normalfont\bfseries}}
\makeatother

\newcommand{\preprintdate}{Month yyyy}
\newcommand{\preprintnumber}{LBNL-42573\\
                             OUTP-98-84-P\\
                             SNS/PH/1998-023\\
                             UCB-PTH-98/58}
\newcommand{\hepnumber}{hep-ph/9812239}
\newcommand{\titletext}{Fermion masses and symmetry 
breaking of a U(2) flavour symmetry\footnote
{This work was supported in part by the U.S. 
Department of Energy under Contracts DE-AC03-76SF00098, in part by the 
National Science Foundation under grant PHY-95-14797
and  in part by the TMR Network under the EEC Contract
No. ERBFMRX - CT960090. }}
\newcommand{\authortext}{
\textbf{\smallskip Riccardo Barbieri$^{\, a}$,
                 Leonardo Giusti$^{\, a}$,\\
                 Lawrence J. Hall$^{\, b}$,
                 Andrea Romanino$^{\, c}$}
\medskip\\
\em\normalsize 
$\mbox{}^a$ Scuola Normale Superiore
and INFN, Sezione di Pisa, I-56126 Pisa, Italy
\\[0.1\baselineskip] 
$\mbox{}^b$ Department of Physics and
Lawrence Berkeley National Laboratory
University of California, Berkeley, California 94720
\\[0.1\baselineskip] 
$\mbox{}^c$ Department of Physics, Theoretical Physics, University of Oxford,
Oxford OX1 3NP, UK}
\newcommand{\abstracttext}{We show how a specific sequential breaking pattern 
of a U(2) flavour symmetry
occurs automatically in a broad framework. The relative orientation in U(2)
space of the spurion fields that breaks the U(2) symmetry is uniquely fixed,
thus determining the form of the fermion mass matrices in a predictive way.}

\newcommand{\makeonlyfirstpage}{%
\setlength{\topmargin}{21mm}
\setlength{\textheight}{172mm}
\renewcommand{\footnoterule}{} 
\title{%
\normalsize\hspace*{\fill}
\begin{tabular}{l}\preprintnumber\\\hepnumber\end{tabular}
\vspace{3\baselineskip}\\\huge\bfseries\titletext}  
\author{
\begin{minipage}[t]{0.8\textwidth}
\large\centering\authortext
\end{minipage}}
\date{\preprintdate}
\begin{document}
\maketitle
\thispagestyle{empty}
\begin{abstract}\large\noindent\abstracttext\end{abstract}
\end{document}}

\title{%
\normalsize\hspace*{\fill}
\begin{tabular}{l}\preprintnumber\\
\end{tabular}
\vspace{3\baselineskip}\\\LARGE\bfseries\titletext}
\author{\begin{minipage}[t]{0.8\textwidth}
\large\centering\authortext
\end{minipage}}
\date{}

\begin{document}
\bigskip
\maketitle
\begin{abstract}\normalsize\noindent\abstracttext
\end{abstract}
\normalsize\vspace{\baselineskip}

\noindent
\section{Introduction and main results}\label{sect:intro}
In previous papers \cite{list}, some of us have pointed out that  a U(2) symmetry might be
relevant to understand several features of flavour physics. We have in
mind both a qualitative and partly quantitative explanation of the pattern
of fermion masses and mixings as well as a possible understanding of
the Flavour Changing Neutral Current problem in supersymmetric
extensions of the Standard Model \cite{pomarol,list}. Along these lines we want to discuss
in this paper the problem of the relative orientation, in U(2) space,
of the fields that break the U(2) symmetry, thus determining in a
unique way the form of the fermion mass matrices. A
refined discussion of the predictions of the U(2) symmetry is given in
Ref.~\cite{BHR}.

The flavour U(2) group acts on the lighter 2 generations $\psi_a$,
$a=1,2$ as a doublet and on the third generation $\psi_3$, like on the
Higgs fields, $H$, as trivial singlets.  In the limit of unbroken
U(2), only the third generation of fermions can acquire a mass,
whereas the first two generations of scalar superpartners are exactly
degenerate. While the first property is not a bad approximation of the
fermion spectrum, the second one is what one needs to keep under
control FCNC and CP-violating phenomena generated by superparticle
exchanges. Furthermore, a two step breaking pattern of U(2) 
accommodates the double hierarchy $m_3 \gg
m_2 \gg m_1$ among different generations in the fermion spectrum.
Although it is natural to view U(2) as a subgroup of U(3), the maximal
flavour group in the case of full intra-family gauge unification, U(3)
will be anyhow strongly broken to U(2) by the large top Yukawa
coupling.

Since the Higgs bosons are flavour singlets, the Yukawa
interactions transform under U(2) as: $(\psi_3 \psi_3)$, $(\psi_3 \psi_a)$,
$(\psi_a \psi_b)$.  Hence the only relevant U(2) representations for
the fermion mass matrices are $1$, $\phi^a$, $S^{ab}$ and $A^{ab}$,
where $S$ and $A$ are symmetric and antisymmetric tensors, and the
upper indices denote a U(1) charge opposite to that of
$\psi_a$. 
We view $\phi^a, S^{ab}$ and
$A^{ab}$ as ``effective flavon'' fields---in general they are polynomials
of the fundamental flavon fields of the theory. The Yukawa potential
has the form 
\begin{equation}
\label{hYH}
{\cal L }_Y = H \left( \psi_3\psi_3 + \psi_3\frac{\phi^a}{M}\psi_a +
\psi_a\frac{S^{ab}}{M}\psi_b 
+ \psi_a\frac{A^{ab}}{M}\psi_b \right)\; ,
\end{equation}
where $M$ is a mass scale weighting all non-renormalizable
interactions and intra-family (vertical) indices and dimensionless
couplings are omitted.

The most general form for the vacuum expectation values
(vevs) of the flavon fields is, in a suitable basis,
\begin{equation}
\label{Avv}
\Vev{A}=
\begin{pmatrix}
0 & v \\
-v & 0
\end{pmatrix}
\quad \Vev{\phi}=
\begin{pmatrix}
0 \\
V
\end{pmatrix}
\quad \Vev{S}=
\begin{pmatrix}
s_2 & s_1 \\
s_1 & s_0
\end{pmatrix}
\end{equation}
with $V,v$ real and positive and $s_i$ in general complex. Let us consider 
first the
possible breaking patterns due to the flavon fields $A$ and
$\phi$ only, leaving $S$ apart for the moment. With $\Vev{A}$ and $\Vev{\phi}$
there are only two ways of breaking U(2) depending on which one of the
two scales $V$ or $v$ is larger:
\begin{align}
\label{ifV}
\text{{\itshape i)} if $V>v$} \quad & U(2)
\overset{V}{\underset{\Vev{\phi}}{\longrightarrow}} U(1)
\overset{v}{\underset{\Vev{A}}{\longrightarrow}} \{e\}\; , \\
\text{{\itshape ii)} if $V<v$} \quad & U(2)
\overset{v}{\underset{\Vev{A}}{\longrightarrow}} SU(2)
\overset{V}{\underset{\Vev{\phi}}{\longrightarrow}} \{e\},
\end{align}
where U(1) corresponds, in the chosen basis, to the subgroup of phase
rotations of the first generation and $e$ is the unity of U(2).
Since {\itshape ii)} would give approximately equal masses for the
first two generations of fermions, the phenomenology selects case
{\itshape i)} and indicates that $V\gg v$. 

Let us now
consider the flavon $S$. To preserve this breaking pattern 
requires $s_0\equiv\Vev{S^{22}}$, which breaks U(2) to the same U(1) as
$\Vev{\phi}$, at the scale $V$ or less and $s_1$ and $s_2$,
which break the remaining U(1), at the scale $v$ or
less. Otherwise, the hierarchy between the first two generations
associated with the ratio $v/V$ could be spoiled.
Furthermore, in order to preserve the phenomenologically successful
relations \cite{bib4,bib5}
\globallabel{eq:Vus}   
\begin{align}  
|V_{us}|  &=  \left| \sqrt{\frac{m_d}{m_s}} + e^{i\Phi}
\sqrt{\frac{m_u}{m_c}} \right|,\mytag \\
\fracwithdelims{|}{|}{V_{ub}}{V_{cb}}  &= 
\sqrt{\frac{m_u}{m_c}},\mytag\\
\fracwithdelims{|}{|}{V_{td}}{V_{ts}}  &=  \sqrt{\frac{m_d}{m_s}}\mytag
\end{align}
among the CKM matrix elements, the masses of light quarks
and the CP-violating phase $\Phi$, stronger constraints on $s_1$, $s_2$ must be
fulfilled \cite{HR,list}:
\begin{equation}
\label{s1v}
|s_1|\ll v, \quad |s_2|\ll \frac{v^2}{V}.
\end{equation}
Should relations~(\ref{s1v}) be considered an ``ad hoc'' hypothesis or
is it possible to justify them?
 
This paper answers this question for
theories having fundamental flavons $\phi, S, A$, their U(2) conjugates
$\bar{\phi}, \bar{S}, \bar{A}$ and possibly U(2) singlets $X_i$. In
these theories~(\ref{s1v}) is a prediction, following from the most
general softly 
broken supersymmetric potential (including possible non-renormalizable
terms) which yields the breaking pattern~(\ref{ifV}) with $V \gg
v$. This result adds confidence to the U(2) scheme and strengthens
its predictions. More precisely, for generic values
of the parameters in the potential of the flavon fields, we show in
Sect.s~\ref{proof} and \ref{full} that the minimum is non degenerate
and that, at the minimum, in an appropriate U(2)-basis, 
\begin{equation}
\frac{\phi}{M}=\ord{
\begin{matrix} \epsilon\epsilon' \\ \epsilon
\end{matrix}} \quad
\frac{S}{M}=\ord{
\begin{matrix} {\epsilon'}^2 & \epsilon\epsilon' \\ \epsilon\epsilon'
& \epsilon 
\end{matrix}} \quad \epsilon\equiv
\frac{V}{M} \quad \epsilon' \equiv \frac{v}{M}, 
\label{pMO}
\end{equation}
with $\bar{\phi}$, $\bar{S}$ having similar magnitude and orientation
as $\phi$ and $S$ respectively. 

By inserting~(\ref{pMO}) in~(\ref{hYH}), we find
the Yukawa matrices in flavour space for the quarks and charged
leptons with the texture
\begin{equation}
\label{eee}
\lambda = 
\begin{pmatrix}
\ll\epsilon^{'2}/\epsilon & \epsilon' & \ll \epsilon' \\
-\epsilon' & \epsilon & {\cal O}(\epsilon) \\
\ll \epsilon' & {\cal O}(\epsilon) & {\cal O}(1)
\end{pmatrix}\; ,
\end{equation}
up to irrelevant phase factors.
The parameters $\epsilon,\epsilon'$ will depend in general upon the fermion
charge. As shown in \cite{HR,list} this texture for quarks leads to the relations
(\ref{eq:Vus}\sep{a}-\sep{c}) as well as to the qualitative relation 
$|V_{cb}|\sim m_s/m_b$.
The phenomenological
consequences of~(\ref{eee}) are carefully analysed in Ref.~\cite{BHR}.

\section{Minimizing the potential in the supersymmetric limit}
\label{proof}
Since we work in a supersymmetric framework, the potential $V$ 
consists of a supersymmetry conserving and a supersymmetry breaking
piece
\begin{equation}
\label{VpS}
V\left(\phi,S,A,\bar{\phi},\bar{S},\bar{A};X_i
\right) = V^{\text{susy}} + V^{\text{breaking}}
\end{equation}
both invariant under a global U(2) symmetry. 
$V^{\text{susy}}$ is determined by a superpotential
$W$ and is characterized by a scale $M$, e.g.\ the GUT scale, much
bigger than the scale $m$, which controls
the size of $V^{\text{breaking}}$ in the usual way and is 
of the order of the electroweak scale.  

Let us consider first the supersymmetric limit. Neglecting non
renormalizable terms, the most general $W$, after a rescaling of the
fields, is
\begin{equation}
  \label{WRl}
  W^{(r)} = \phi\bar{S}\phi + \bar{\phi}S\bar{\phi} +X_\phi \ppb+ X_S\bar{S}S +
  X_A\bar{A}A.
\end{equation}
$X_\phi$, $X_S$ and $X_A$ are linear combinations of one or more singlet fields
and of possible mass terms.
The part of the
superpotential only dependent on the singlet fields $X$'s does not affect any
of our considerations and it is therefore not explicitly shown. Couplings 
between $A$, $\bar{A}$ and the other fields are forbidden by
the antisymmetry of $A$, $\bar{A}$ in the $U(2)$ indices. 

If $X_S\neq 0$, there
is a supersymmetric minimum where
\begin{equation}
  \label{Sab}
  \left\{
    \begin{aligned}
      S^{ab} &= -\frac{\phi^a\phi^b}{X_S} \\
      \bar{S}_{ab} &=
      -\frac{\bar{\phi}_a\bar{\phi}_b}{X_S} 
    \end{aligned}
  \right.
\end{equation}
and 
\begin{equation}
  \label{ppX}
\ppb=\frac{X_S X_\phi}{2}.
\end{equation}
To show that this solution is preferable to $\phi=\bar{\phi}=0$ 
supersymmetry breaking must be considered, as done in Section 
\ref{full}. 

The minimum equations for $A$, $\bar{A}$ 
\begin{equation}
  \label{XAA}
  X_A A^{ab} = 0 \quad X_A \bar{A}_{ab} = 0
\end{equation}
decouple, at renormalizable level. For $X_A\neq 0$, they give
$A=\bar{A}=0$. Unlike the case for $S$, $\bar{S}$, $\phi$,
$\bar{\phi}$, the introduction of non-renormalizable interactions
cannot be treated pertubatively.

Let us therefore consider first, in the case of a general
superpotential, the minimum equations for $S$, $\bar{S}$. For field
vevs small relative to $M$, standard inversion theorems guarantee
that the minimum equations can be solved for
$S$, $\bar{S}$ functions of $\phi$, $\bar{\phi}$ as for eq.~(\ref{Sab}). 
>From general U(2)
covariance,
\globallabel{Sab2}
\begin{align}
  S^{ab} &= \Sigma_1 \phi^a\phi^b + \Sigma_2 \left[(A\bar{\phi})^a\phi^b + \phi^a
  (A\bar{\phi})^b\right] + \Sigma_3 (A\bar{\phi})^a(A\bar{\phi})^b 
\equiv \hat{S}^{ab}\mytag
  \\
  \bar{S}_{ab} &= \bar{\Sigma}_1 \bar{\phi}_a\bar{\phi}_b + \bar{\Sigma}_2
  \left[(\bar{A}\phi)_a\bar{\phi}_b + \bar{\phi}_a 
  (\bar{A}\phi)_b\right] + \bar{\Sigma}_3 (\bar{A}\phi)_a(\bar{A}\phi)_b
\equiv \hat{\bar{S}}_{ab}, \mytag
\end{align}
with $\Sigma_i$, $\bar{\Sigma}_i$ functions of the invariants $\ppb$,
$\aab$. More explicitly, to leading order in $1/M$,
\globallabel{Sab3}
\begin{align}
  S^{ab} &= \sigma_1\frac{\phi^a\phi^b}{X_S} + \frac{\sigma_2}{MX_S}
  \left[(A\bar{\phi})^a\phi^b + \phi^a 
  (A\bar{\phi})^b\right] + \frac{\sigma_3}{MX_S^2}
  (A\bar{\phi})^a(A\bar{\phi})^b \mytag 
  \\
  \bar{S}_{ab} &= \bar{\sigma}_1\frac{\bar{\phi}_a\bar{\phi}_b}{X_S} +
  \frac{\bar{\sigma}_2}{MX_S} 
  \left[(\bar{A}\phi)_a\bar{\phi}_b + \bar{\phi}_a 
  (\bar{A}\phi)_b\right] + \frac{\bar{\sigma}_3}{MX_S^2}
  (\bar{A}\phi)_a(\bar{A}\phi)_b, \mytag 
\end{align}
where $\sigma_1=\bar{\sigma}_1=-1$ and
$\sigma_{2,3}$, $\bar{\sigma}_{2,3}$ are polynomial in $\ppb/X_S^2$,
$\aab/X_S^2$, $X_A/X_S$. It is possible to give examples of explicit
non-renormalizable potential that generate non vanishing
$\sigma_{2,3}$, $\bar{\sigma}_{2,3}$. 

To solve the minimum equations $\partial W/\partial\phi = \partial
W/\partial\bar{\phi}=0$, it is useful to define
\begin{equation}
  \label{Wpp}
  \hat{W}(\ppb,\aab)\equiv 
  W\left( \hat{S}(\phi,\bar{\phi},A,\bar{A}),
    \hat{\bar{S}}(\phi,\bar{\phi},A,\bar{A}),
  \phi,\bar{\phi}, A, \bar{A}\right).
\end{equation}
Since $\hat{S}$, $\hat{\bar{S}}$ solve $\partial W/\partial S = \partial
W/\partial\bar{S}=0$, it is immediate that 
\begin{equation}
  \label{WpW2}
  0 = \frac{\partial W}{\partial \bar{\phi}_a} = \frac{\partial
  \hat{W}}{\partial \ppb} \phi^a \quad  
  0 = \frac{\partial W}{\partial \phi^a} = \frac{\partial
  \hat{W}}{\partial \ppb} \bar{\phi}_a
\end{equation}
which, again disregarding the possibility
$\phi=\bar{\phi}=0$, are equivalent to the unique equation
\begin{equation}
  \label{eqP}
  \frac{\partial \hat{W}}{\partial \ppb}=0\; .
\end{equation}
This allows to compute $\ppb$ in terms of $\aab$ and of the singlet
fields. As before, to leading order in $1/M$, from eq.~(\ref{eqP}),
\begin{equation}
  \label{pp1nr}
  \ppb =\frac{X_S}{2}\left(X_\phi
  +\frac{\sigma_\phi(\aab/X_S^2)}{M}\right),
\end{equation}
with $\sigma_\phi$ polynomial in its variable.

Analogously, $\partial \hat{W}/\partial\aab=0$ is the unique equation to be
solved in $\aab$. Examples of non renormalizable interactions that fix $\aab$
at a nonvanishing vev are easy to construct. We assume $|\aab|\ll|\ppb|$. 

These considerations make clear that the supersymmetric minimum
is highly degenerate even for a non renormalizable potential. 
Other than the degeneracy related to U(2)
invariance, the surface of minima is flat in directions corresponding
to the relative orientation of $\phi$ and $\bar{\phi}$ and to the
rescalings 
\begin{equation}\label{eq:rescaling}
\phi\rightarrow x\phi,\quad  \bar{\phi}\rightarrow
\bar{\phi}/x, \quad A\rightarrow yA, \quad \bar{A}\rightarrow 
\bar{A}/y
\end{equation} 
with $x$, $y$ real. This degeneracy is removed by the
introduction of the supersymmetry breaking potential, as we
show in the next Section. Furthermore, if the parameters in the potential related to
fields and bar-fields have similar order of magnitude, we show that
\begin{equation}
\label{p1p}
 \bar{\phi_1}\phi^\dagger_2 - \bar{\phi_2}\phi^\dagger_1 = 
{\cal O} \left(\frac{V^2 v}{M}\right),\quad
 \ppd\sim \bppd,\quad \aad\sim\baad
\end{equation}
which, together with $X_S\sim X_\phi$,
proves the results stated in Sect.~\ref{sect:intro}. This is made manifest by
choosing a basis where $\phi_1=0$, since, from (\ref{p1p}), 
$\bar{\phi_2}/M\simeq\phi_2/M\simeq\epsilon$, 
$\bar{\phi_1}/M\simeq\epsilon\epsilon'$ and $S^{ab}$, 
by inserting these $\phi$, $\bar{\phi}$ vevs in (\ref{Sab3}\sep{a}),
has the same form as in (\ref{pMO}). The proof of (\ref{p1p}) is lengthy. We
establish it by discussing a general method to minimize the potential in
presence of supersymmetry breaking. 

\section{Minimizing the full potential}
\label{full}
Denoting by $z_i$ the 
collection of all fields, the potential $V$
consists of three pieces, $V=V_0+V_1+V_2$, with the general structure 
\globallabel{V0p}
\begin{align}
V_0 &= \frac{\partial W^\dagger}{\partial z_i^\dagger} \frac{\partial
W}{\partial z_i} \mytag \\
V_1 &= m f(z) + m f^\dagger(z) \mytag \\
V_2 &= \sum_i m^2_i |z_i|^2\; , \mytag
\end{align}
and $f(z)$ holomorphic in $z$.
Let us write the position of the minimum of the full potential as $z_i
= z_i^{(0)} + z_i^{(1)}$ where $z^{(0)}$ is on the surface ${\cal S}$
of minima of $V_0$ , as defined by (\ref{eq:rescaling}), and $z^{(1)}$ is a correction, due to the
supersymmetry breaking terms, orthogonal to it. Note that, due to
the structure of the supersymmetric minimum, each holomorphic 
U(2)-invariant function
of the flavon fields $\phi$, $S$, $A$, $\bar{\phi}$, $\bar{S}$,
$\bar{A}$ or of their hermitian conjugates, separately, is constant on
${\cal S}$. Therefore to resolve the degeneracy
of $z^{(0)}$ an expansion of $V$ to first order in $m$ or $z^{(1)}$ is
not sufficient. Expanding $V(z^{(0)} +
z^{(1)})$ up to second order around $z^{(0)}$ gives
\begin{multline}
\label{Vz0}
V(z^{(0)} + z^{(1)}) = \\
c+\left.\frac{\partial^2 V_0}{\partial z_i
\partial z^\dagger_j}\right|_{z^{(0)},{z^{(0)}}^\dagger} z^{(1)}_i
{z^{(1)}}^\dagger_j + m \left.\frac{\partial f}{\partial
z_i}\right|_{z^{(0)}} z_i^{(1)} + m \left.\frac{\partial
f^\dagger}{\partial z^\dagger_i}\right|_{{z^{(0)}}^\dagger}
{z_i^{(1)}}^\dagger + \left.V_2\right|_{z^{(0)},{z^{(0)}}^\dagger},
\end{multline}
where $c=V_0(z^{(0)},{z^{(0)}}^\dagger) + m f(z^{(0)}) + m
f^\dagger({z^{(0)}}^\dagger)=m f(z^{(0)}) + m
f^\dagger({z^{(0)}}^\dagger)$ is actually independent of
$z^{(0)},{z^{(0)}}^\dagger$ and all other terms are of second order in
the supersymmetry breaking scale $m$. 
While the constant $c$ is independent of the position on the surface ${\cal S}$, it
does distinguish this surface from the alternative solution in which
$\phi=\bar{\phi}=0$. In order to select the desired solution, $c$ must be
negative at the minimum. It is easy to convince oneself that this is
the case in a large region of the parameter space. 

Since $z = z^{(0)} + z^{(1)}$ is a minimum, by differentiating~(\ref{Vz0})
with respect to $z^{(1)}$ one obtains
\begin{equation}
\label{nai}
\left.\frac{\partial^2 V_0}{\partial z_i
\partial z^\dagger_j}\right| {z^{(1)}}^\dagger_j + m
\left.\frac{\partial f}{\partial z_i}\right| = 0
\end{equation}
which, substituted into~(\ref{Vz0}), leads to
\begin{equation}
\label{VV2}
V = V_2 + m \frac{\partial f^\dagger}{\partial
z^\dagger_i} {z_i^{(1)}}^\dagger = V_2 + m \frac{\partial
f}{\partial z_i} z_i^{(1)} \equiv V_{\text{eff}},
\end{equation}
where $c$ has been omitted. $V_{\text{eff}}$, with $z^{(1)}$ given
by~(\ref{nai}) in terms of $z^{(0)}$ and proportional to the
holomorphic supersymmetry breaking terms, can be viewed as a
simplified ``effective'' potential to be minimized in $z_i^{(0)}$,
${z_i^{(0)}}^\dagger$ on the surface ${\cal S}$, thus
removing the degeneracy of the supersymmetric minimum.

Notice that the contribution due to the holomorphic supersymmetry
breaking terms vanishes if they are universal, namely if
$f\propto W$. In this case we have in fact $\partial f/\partial
z\propto \partial W/\partial z=0$ on ${\cal S}$.

Let us therefore consider the general problem of the minimization on
${\cal S}$ of a
U(2)-symmetric potential $V_{\text{eff}}$, function of the
flavon fields and their hermitian conjugates. On ${\cal S}$ we can use 
\begin{multline}
\label{Vef}
V_{\text{eff}}\left(S=\hat{S}, \bar{S}=\hat{\bar{S}}, \phi,
\bar{\phi},\text{h.c.}\right) \equiv \\
\hat{V}_{\text{eff}}\left(\ppd,\bppd,\aad,\baad,\pa,\pab,\pad,\pabd\right),
\end{multline}
where the variables of $\hat{V}_{\text{eff}}$ are all the possible
U(2)-invariants one can build with the flavons and their hermitian
conjugates besides $\ppb$, $\dppb$, $\aab$, $\daab$ that are constant
on ${\cal S}$. $\hat{V}_{\text{eff}}$ has to be minimized under the
constraint $\ppb = (\ppb)^{(0)}$, $\dppb =
(\dppb)^{(0)}$, $\aab = (\aab)^{(0)}$, $\daab =
(\daab)^{(0)}$. Introducing the Lagrange multipliers
\begin{multline}
\label{VVe}
\hat{V} \equiv \hat{V}_{\text{eff}} - \lambda_\phi(\ppb-(\ppb)^{(0)}) -
\lambda_\phi^\dagger(\dppb-(\dppb)^{(0)}) - \\
\lambda_A(\aab-(\aab)^{(0)}) - 
\lambda_A^\dagger(\daab-(\daab)^{(0)}),
\end{multline}
and projecting the minimum equation for $\phi$ along two orthogonal
directions one gets 
\globallabel{lm}
\begin{gather}  
0 = (\bar{\phi}^\dagger)^a \frac{\partial\hat{V}}{\partial\phi^a} = 
\frac{\partial\hat{V}_{\text{eff}}}{\partial\ppd}\dppb
-\lambda_\phi\bppd \mytag \\
0 = \phi^\dagger_a A^{ab} \frac{\partial\hat{V}}{\partial\phi^b} = 
\frac{\partial\hat{V}_{\text{eff}}}{\partial\ppd}\pa -\frac{1}{2}\left(
\frac{\partial\hat{V}_{\text{eff}}}{\partial\pab}\bppd\aab -
\frac{\partial\hat{V}_{\text{eff}}}{\partial\pad}\bppd\aad
\right) \mytag \\
0 = \frac{\partial\hat{V}}{\partial A^{ab}} A^{ab} =
\frac{\partial\hat{V}_{\text{eff}}}{\partial\pa}\pa +
\frac{\partial\hat{V}_{\text{eff}}}{\partial\aad}\aad. \mytag 
\end{gather}
Similarly one gets other 9 barred and hermitian
conjugate equations. The lagrange multipliers can be eliminated from
this total of 12 equations leaving only the following 4 independent
(``real'') equations:
\globallabel{4eq}
\begin{gather}
\frac{\partial\hat{V}_{\text{eff}}}{\partial\ppd}\ppd =
\frac{\partial\hat{V}_{\text{eff}}}{\partial\bppd}\bppd \mytag \\
\frac{\partial\hat{V}_{\text{eff}}}{\partial\aad}\aad +
\frac{\partial\hat{V}_{\text{eff}}}{\partial\pa}\pa =  
\frac{\partial\hat{V}_{\text{eff}}}{\partial\baad}\baad +
\frac{\partial\hat{V}_{\text{eff}}}{\partial\pab}\pab \mytag \\
\frac{\partial\hat{V}_{\text{eff}}}{\partial\bppd} \frac{\pa}{\ppd} =
\frac{1}{2}\left(
\frac{\partial\hat{V}_{\text{eff}}}{\partial\pab}\aab -
\frac{\partial\hat{V}_{\text{eff}}}{\partial\pad}\aad 
\right) \mytag \\
\frac{\partial\hat{V}_{\text{eff}}}{\partial\bppd} \frac{\pad}{\ppd} =
\frac{1}{2}\left(
\frac{\partial\hat{V}_{\text{eff}}}{\partial\pabd}\daab -
\frac{\partial\hat{V}_{\text{eff}}}{\partial\pa}\daab 
\right). \mytag
\end{gather}
The number of degree of freedom is 4 (real) too, because all the
U(2)-invariants can be expressed on ${\cal S}$ in terms of $\ppd$,
$\aad$, $\pa$, $\pad$ through
\begin{gather*}
\baad = \frac{(\aab)^{(0)}(\daab)^{(0)}}{\aad}, \quad 
\bppd = \frac{(\ppb)^{(0)}(\dppb)^{(0)}}{\ppd} +2
\frac{(\pabd)(\pa)}{\ppd\aad}, \\
\pab = \pad\frac{(\aab)^{(0)}}{\aad}, \quad 
\pabd = \pa\frac{(\daab)^{(0)}}{\aad}.
\end{gather*}

Notice that the same 4 equations~(\ref{4eq}) could be recovered by
using the previous relations to parametrize the surface ${\cal S}$ and
to express $\hat{V}_{\text{eff}}$ in terms of $\ppd$, $\aad$, $\pa$,
$\pad$ and by differentiating $\hat{V}_{\text{eff}}$ with respect to
them.

Let us consider now as an example of application of the previous
formalism the simple renormalizable case in which the holomorphic
supersymmetry breaking terms are proportional to the
superpotential. In this case $V_{\text{eff}}$ is of the form
\begin{equation}
\label{VeV2}
V_{\text{eff}} = V_2 = m^2_S S^\dagger S + m^2_{\bar{S}}
\bar{S}^\dagger\bar{S} + m^2_\phi \ppd + m^2_{\bar{\phi}} \bppd +
m^2_A A^\dagger A + m^2_{\bar{A}}
\bar{A}^\dagger\bar{A}
\end{equation}
and therefore
\begin{equation}
\label{VeV3}
\hat{V}_{\text{eff}} = m^2_S \frac{(\ppd)^2}{X_S^\dagger X_S} +
m^2_{\bar{S}} \frac{(\bppd)^2}{X_S^\dagger X_S} + 
m^2_\phi \ppd + m^2_{\bar{\phi}} \bppd +
m^2_A A^\dagger A + m^2_{\bar{A}}
\bar{A}^\dagger\bar{A}.
\end{equation}
The equations~(\ref{4eq}) simplify to
\globallabel{lm1}
\begin{gather}
2 m^2_S \frac{(\ppd)^2}{X_S^\dagger X_S} + m^2_\phi \ppd =
2 m^2_{\bar{S}} \frac{(\bppd)^2}{X_S^\dagger X_S} + m^2_{\bar{\phi}} \bppd
\mytag \\
m^2_A A^\dagger A = m^2_{\bar{A}}\bar{A}^\dagger\bar{A} \mytag \\
\pad = 0 \mytag \\
\pa = 0. \mytag
\end{gather}
Since $\pa=0$, $\bar{\phi}$ and $\phi^\dagger$ are aligned and
$\ppd = x X_S^\dagger X_S/2$, $\bppd = \bar{x} X_S^\dagger X_S/2$ with
$x\bar{x}=1$ and $x$ determined by eq.~(\ref{lm1}\sep{a}) provided that
a positive solution exists. Therefore
$\ppd$ and $\bppd$ are of the same order of magnitude if $m^2_S\sim
m^2_{\bar{S}}$ and $m^2_\phi\sim m^2_{\bar{\phi}}$. Analogously for
$\aad$ and $\baad$. 

The more general renormalizable case in which the $A$-terms are
generic can be solved more easily by using a symbolic manipulation
program. However, more than the explicit form of the equations the
important outcomes are that: {\itshape i)} in a large region of the
parameter space, eqs.~(\ref{4eq}\sep{a,}\sep{b}) have a solution, the
degeneracy is then removed and for similar values of parameters and
``barred'' parameters $|\ppd|\sim|\bppd|\sim V^2$, $|\aad|\sim|\baad|\sim
v^2$;  {\itshape ii)} $V_{\text{eff}}$ does not depend on $\pa$, as it
can be easily seen. Point {\itshape i)} assures that, in a neighborhood of the
$1/M=0$ case, the general non-renormalizable case can be solved
pertubatively from the renormalizable one, thereby resolving the
degeneracy. Point {\itshape ii)},
together with eqs.~(\ref{4eq}\sep{c,}\sep{d}), assures that $\pa=0$ and
therefore that $\phi$ and $\bar{\phi}$ are aligned in the
renormalizable case. Moreover, from perturbative expansion one gets 
$|\pa|\sim v^2 V^2/M$
and therefore 
\[
|\bar{\phi}_1|\sim \frac{v V}{M}
\]
in the basis in which $\phi^1=0$. This completes the proof of (\ref{p1p}).


\end{document}